\documentclass[12pt]{article}

\usepackage{epsfig}
\usepackage{epstopdf}
\usepackage{amssymb,amsmath}
\usepackage{comment}

\usepackage[sorting=none,citestyle=numeric-comp]{biblatex}
\usepackage{hyperref}

\newcommand{\bea}{\begin{eqnarray}}
\newcommand{\eea}{\end{eqnarray}}

\numberwithin{equation}{section}

\begin{document}

\begin{titlepage}
%
%
\vspace*{10mm}
\begin{center}
\baselineskip 25pt 
{\Large\bf
Double Inflation in Classically Conformal $B-L$ Model 
}
\end{center}
\vspace{5mm}

\begin{center}
{\it
Anish Ghoshal$^{~a,}$\footnote{anish.ghoshal@fuw.edu.pl}, 
Nobuchika Okada$^{~b,}$\footnote{okadan@ua.edu}, 
Arnab Paul$^{~c,}$\footnote{arnabpaul9292@gmail.com}
and Digesh Raut$^{~d,}$\footnote{draut@smcm.edu}
}
\end{center}
\begin{center}
{\it
$^{a}$ Institute of Theoretical Physics, Faculty of Physics, \\ University of Warsaw, ul. Pasteura 5, 02-093 Warsaw, Poland, \\ 
$^{b}$ University of Alabama, Tuscaloosa, Alabama 35487, USA \\
$^{c}$ Centre for Strings, Gravitation and Cosmology, Department of Physics, Indian Institute of Technology Madras, Chennai 600036, India\\
$^{d}$ Department of Physics and Materials Science, St. Mary's College of Maryland, 
St. Mary's City, Maryland 20686, USA
}
\end{center}

\vspace{0.5cm}
\begin{abstract}
It has recently been shown in Ref.~\cite{Karam:2023haj} that the double-inflation scenario based on the Coleman-Weinberg potential can successfully generate primordial black holes (PBHs) with the inflationary predictions consistent with the Planck measurements. These PBHs can play the role of dark matter in our universe. 
In this paper, we propose the classically conformal minimal $B-L$ model as an ultra-violet (UV) completion of the scenario. 
In our model, the $B-L$ Higgs field is identified with the inflaton and the electroweak symmetry breaking is triggered by the radiative $B-L$ symmetry breaking with the Coleman-Weinberg potential. 
We show that this UV completion leads to a viable cosmological history after the double inflaton: the universe is reheated via inflaton decay into right-handed neutrinos whose mass is determined consistently by a relation between the number of e-folds and reheating temperature. 
Using the general parameterization for neutrino Dirac Yukawa couplings through the seesaw mechanism and the neutrino oscillation data, we also show that the observed baryon asymmetry of the universe is successfully reproduced by either resonant leptogenesis or non-thermal leptogenesis.
Based on the scalar power spectrum shown in Ref.~\cite{Karam:2023haj}, we evaluate the scalar induced gravitational wave spectrum, which can be tested by various proposed gravitational wave observatories like BBO, DECIGO etc.
\end{abstract}
\end{titlepage}

\section{Introduction}

Cosmic inflation, which is the accelerated expansion of the universe at early time,  solves the horizon and the flatness problem as well as provides the primordial density fluctuations essential for the large scale structure formation \cite{Starobinsky:1980te,Guth:1980zm,Sato:1980yn,1980ApJ...241L..59K}. 
Rapidly increasing precision and cosmological data from various observations, in particular, Cosmic Microwave Background (CMB) measurement \cite{Akrami:2018odb} has provided severe constraints on inflationary potential.

The inflaton field generates density fluctuation not only at the CMB scale, $ {\rm  k_{CMB}/{\rm Mpc^{-1}}}  \sim 10^{-2} -10^{-1}$, but also over a wide range of scales with ${\rm k\gg k_{CMB}}$. 
If the density fluctuations at some scale are significantly enhanced and the scalar power spectrum $P_\zeta  \gtrsim {\cal O } (10^{-3})$, which is much greater than the power spectrum $P_\zeta  \gtrsim {\cal O } (10^{-9})$ corresponding to the CMB scales, primordial black hole (PBH) can be formed. It is particularly interesting that such PBHs can be a dark matter candidate in our Universe. See \cite{Carr:2020gox} for a recent review.

In the context of a single-field inflation scenario, it has been shown that the scalar power spectrum can be enhanced if the potential exhibits an approximate inflection point. If the inflaton passes through this region with a suitably small velocity, termed as the ultra-slow roll, the density perturbations can be significantly enhanced to form PBHs \cite{Kawasaki:1997ju, Byrnes:2018txb, Carrilho:2019oqg,Ragavendra:2020sop, Cole:2022xqc, Ghoshal:2023wri, Ghoshal:2023pcx}. 
See \cite{Sasaki:2016jop, Escriva:2022duf,Ragavendra:2023ret} for a recent review. 
Such an enhancement generally requires fine-tuning of model parameters, with which the predicted value of the spectral index mostly does not match the CMB data \cite{Akrami:2018odb}.
Successful PBH formation has also been discussed in the multi-field inflation scenarios \cite{Braglia:2020eai,Raveendran:2022dtb}, in hybrid inflation \cite{Clesse:2015wea, Afzal:2024xci}, due to dynamics of spectator fields \cite{Chen:2023lou} or others \cite{Ashoorioon:2019xqc,Heydari:2021gea,Kawai:2022emp,Kawai:2021edk}.

Another possibility for the single field inflation is the so-called double inflation scenario \cite{Karam:2023haj} in which the inflaton has a double-well potential. 
The first inflation takes place when the inflaton rolls down from a large field value, which produces density fluctuations responsible for the CMB spectrum. After the first inflation, the field passes through a potential minimum and climbs up the potential hill towards the potential origin. When the inflaton almost comes to a stop in the vicinity of the potential origin, the second inflation occurs. 
If the inflaton potential is very flat near the origin, the ultra-slow role condition is satisfied to enhance the density fluctuation during the second inflation, enough to form the PBHs.

Recently Ref.~\cite{Karam:2023haj} analyzed the double-inflation scenario based on the Coleman-Weinberg potential \cite{Coleman:1973jx} to show that (i) the inflationary predictions from the first inflation are consistent with the Planck data \cite{Akrami:2018odb} and (ii) the density fluctuation is sufficiently enhanced during the second inflation for production of PBH. Note that the Coleman-Weinberg potential is ideal for realizing this scenario since the mass-squared (the second derivative of the potential) of the field at the origin is zero. In this paper, we propose classically conformal $U(1)$ $B-L$ (Baryon number minus Lepton number) extended Standard Model (SM) as an ultra-violet (UV) completion of the double inflation scenario analyzed in Ref.~\cite{Karam:2023haj}.

From the particle physics viewpoint, the classically conformal $U(1)_{B-L}$ extended SM \cite{Iso:2009ss,Iso:2009nw} is particularly interesting since it can address several outstanding problems in the SM. 
The $B-L$ symmetry is the unique anomaly-free global symmetry in the SM. 
The so-called minimal $B-L$ model \cite{Davidson:1978pm,Mohapatra:1980qe,Marshak:1979fm,Wetterich:1981bx,Masiero:1982fi,Mohapatra:1982xz} is the minimal framework for extending the SM to gauge the $B-L$ symmetry. 
The particle content of the SM is extended to include three right-handed neutrinos to cancel $B-L$ related gauge anomalies and the $B-L$ Higgs field to break the $B-L$ symmetry. 
Since the $B-L$ symmetry breaking generates Majorana masses for the right-handed neutrinos, the Type-I seesaw mechanism \cite{Minkowski:1977sc,Yanagida,GRS,Glashow,Mohapatra:1979ia,Schechter:1980gr} is implemented to explain the origin of the light neutrino masses. 
The model can also explain the origin of baryon-asymmetry of the universe through leptogenesis\cite{Fukugita:1986hr}. 
We consider the classically conformal extension of the minimal $B-L$ model \cite{Iso:2009ss,Iso:2009nw}\footnote{See Refs.~\cite{Barman:2021lot,Barman:2022njh,Dasgupta:2022isg,Ghoshal:2022hyc} for various cosmological implications of conformal B-L scenarios.}, where the $B-L$ gauge symmetry is radiatively broken by the Coleman-Weinberg mechanism. 
This also triggers the breaking of the electroweak symmetry, and therefore, the electroweak symmetry breaking originates from the radiative $B-L$ symmetry breaking.

In the framework of the classically conformal $B-L$ model, we identify the $B-L$ Higgs field as the inflaton in the double-inflation scenario \cite{Karam:2023haj}. 
Focusing on a benchmark parameter set to realize a successful double inflation, we show the successful thermal history of the universe after the double inflation, namely, reheating after the double inflation, leptogenesis, electroweak symmetry breaking and neutrino mass generation. 
We also show that enhanced density fluctuations needed for the PBH production produce (scalar-induced) gravitational waves, which can be tested by various planned observations in the future \cite{Domenech:2021ztg}.

\textit{The paper is organized as follows:} in section 2, we introduce classically conformal $B-L$ model with radiative symmetry breaking. In section 3, we discuss the double inflation scenario with the $B-L$ Higgs field identified as inflation. Successful reheating after the double inflation and thermal leptogenesis are discussed in Section 4. Section 5 is devoted to conclusions and discussions. 
In this section, we also discuss the scalar-induced gravitational wave spectrum associated with the enhanced power spectrum to form the PBHs.

\section{Classically Conformal Minimal \texorpdfstring{$B-L$}{BL}  Model}

\begin{table}[h]
\begin{center}
\begin{tabular}{c|ccc|c}
            & SU(3)$_c$ & SU(2)$_L$ & U(1)$_Y$ & U(1)$_{B-L}$  \\
\hline
$ q_L^i $    & {\bf 3}   & {\bf 2}& $+1/6$ & $+1/3$  \\ 
$ u_R^i $    & {\bf 3} & {\bf 1}& $+2/3$ & $+1/3$  \\ 
$ d_R^i $    & {\bf 3} & {\bf 1}& $-1/3$ & $+1/3$  \\ 
\hline
$ \ell^i_L$    & {\bf 1} & {\bf 2}& $-1/2$ & $-1$  \\ 
$ N_R^i$   & {\bf 1} & {\bf 1}& $ 0$   & $-1$  \\ 
$ e_R^i  $   & {\bf 1} & {\bf 1}& $-1$   & $-1$  \\ 
\hline 
$ H$         & {\bf 1} & {\bf 2}& $-1/2$  &  $ 0$  \\ 
$ \Phi$      & {\bf 1} & {\bf 1}& $  0$  &  $+2$  \\ 
\end{tabular}
\end{center}
\caption{Particle content of the minimal $B-L$ extended SM. Besides the SM particles, it includes three right-handed neutrinos $N_R^i$ (where $i=1,2,3$ denotes the generation index) and a complex scalar $\Phi$. }
\end{table}

The particle content of the minimal anomaly-free extension of the SM with gauged $B-L$ (baryon minus lepton number) symmetry is listed in Table~1, where $N_R^i$ ($i=1,2,3$) are three generations of right-hand neutrinos (RHNs), $\Phi$ is the U(1)$_{B-L}$ Higgs field, and the rest are the SM particles.

The Yukawa sector involving the new particles is given by  
\bea  
   {\cal L} \supset   \left(- \sum_{i,j=1}^{3} Y_D^{ij} 
   \overline{N_R^i} \; H^\dagger \ell_L^j  - \frac{1}{2} \sum_{i=1}^{3} Y_M^i  \Phi  \overline{N_{R}^{i~C}} N^i_{R} \right) +{\rm h.c.},
   \label{eq:seesaw0}
\eea 
where the first and the second terms are, respectively, the Dirac and Majorana Yukawa couplings, and we have chosen the flavor diagonal basis for the Majorana Yukawa couplings. 

The renormalizable scalar potential consistent with classically conformal invariance does not include the mass-squared terms at the tree level and is expressed as  
\bea  
V = \lambda_H \left(  H^{\dagger}H  \right)^2
+ \lambda_{\Phi} \left(  \Phi^{\dagger} \Phi   \right)^2
- \lambda_{\rm mix} 
\left(  H^{\dagger}H   \right) 
\left(  \Phi^{\dagger} \Phi  \right) , 
\label{Higgs_Potential}
\eea
where the couplings $\lambda_H,~\lambda_{\Phi}$ and $\lambda_{\rm mix}$ are taken to be positive. 
Assuming $\lambda_{\rm mix}$ is negligibly small, we can analyze the Higgs potentials for $\Phi$ and $H$ separately as a good approximation. This assumption will be justified later. 
Let us first analyze the $B-L$ Higgs sector. 
By choosing the renormalization scale to be the vacuum expectation value (VEV) of $\Phi$, the Coleman-Weinbeg potential \cite{Coleman:1973jx} at the one-loop level is given by 
\begin{eqnarray}
   V(\phi) =  \frac{\lambda_\Phi}{4} \phi^4 
     + \frac{\beta_\Phi}{8} \phi^4 \left(  \ln \left[ \frac{\phi^2}{v_\phi^2} \right] - \frac{25}{6} \right) + V_0, 
\label{eq:CW_potential} 
\end{eqnarray}
where $\phi / \sqrt{2} = \Re[\Phi]$ is a real scalar, $v_\phi = \langle \phi \rangle $ is the VEV of $\Phi$, and  $V_0 = \beta_\Phi v_\phi^4/16$ ensures $V(v_\phi) = 0$. Imposing the stationary condition $\left. dV/d\phi\right|_{\phi=v_\phi} = 0$ we get a relation between the renormalized self-coupling,  
\begin{eqnarray}
   \lambda_\Phi = \frac{11}{6} \beta_\Phi, 
\label{eq:stationary}
\end{eqnarray} 
where 
\begin{eqnarray}
 \lambda_\Phi = \frac{1}{3 !}\left. \frac{d^4V(\phi)}{d \phi^4} \right|_{\phi=v_\phi}, 
\end{eqnarray}  
and the coefficient of the one-loop corrections,  
\begin{eqnarray}
 \beta_\Phi	= \frac{1}{16 \pi^2}  \left( 20\lambda_\Phi^2 + 96 g^4  -  \sum_{i=1}^{3} (Y_M^i)^4 \right) 
	\simeq  
	\frac{1}{16 \pi^2}  \left( 96 g^4  - \sum_{i=1}^{3} (Y_M^i)^4 \right).  
\end{eqnarray}
Here, $g$ is the $B-L$ gauge coupling, and we have used $\lambda_\Phi^2 \ll g^4$ indicated by Eq.~(\ref{eq:stationary}).

The mass of the $B-L$ Higgs boson is given by 
\begin{eqnarray}
  m_\phi^2 &=& \left. \frac{d^2 V}{d\phi^2}\right|_{\phi=v_\phi}  
                   =\beta_\Phi v_\phi^2  \simeq 
  \frac{1}{16 \pi^2} \left( 96 g^4 -  \sum_{i=1}^{3} (Y_M^i)^4 \right) v_\phi^2 \nonumber \\
  &=&   \frac{3}{2\pi^2} g^2 m_{Z^\prime}^2 
     \left( 1-\frac{2}{3} \sum_{i=1}^{3} \left( \frac{m_{N^i}}{m_{Z^\prime}}\right)^4   \right), 
\label{Eq:mass_phi}
\end{eqnarray}
where 
\begin{eqnarray}
  m_{Z^\prime}  = 2 g v_\phi, 
   \;  \;  m_{N^i} = \frac{Y_M^i}{\sqrt 2} v_\phi, 
\label{Eq:mass_Zp_DM}
\end{eqnarray} 
are the $B-L$ gauge boson ($Z^\prime$ boson) and the Majorana RHN masses that are generated by the breaking of $B-L$ symmetry. The vacuum stability condition $m_\phi^2 >0$ demands $m_{Z^\prime} >  \left(\frac{2}{3}\right)^{1/4} \sum_{i=1}^{3} m_{N^i}$, or equivalently, $\sum_{i=1}^{3} Y_M^i < (96)^{1/4} g$.

After the radiative breaking of the $B-L$ symmetry via the Coleman-Weinberg mechanism, the mixing quartic coupling term in Eq.~(\ref{Higgs_Potential})  generates the negative mass squared term for the SM Higgs doublet,  
\begin{equation}
  V \supset  \frac{\lambda_H}{4}h^4 - \frac{\lambda_{\rm mix}}{4} v_\phi^2 h^2,  
\end{equation}
where we have replaced $H$ by $H = 1/\sqrt{2}\, (0 \; \, h)^T$ in the unitary gauge, so that the electroweak symmetry breaking is triggered.  
The SM Higgs boson mass ($m_h$) is expressed as 
\begin{equation}
  m_h^2  = \lambda_{\rm mix} v_\phi^2 = 2 \lambda_H v_h^2, 
\end{equation}
 where $v_h=246$ GeV is the SM Higgs VEV. 
Considering the Higgs boson mass of $m_h=125$ GeV and the LEP experimental constraint on the VEV $v_\phi \gtrsim 10$ TeV \cite{LEP:2003aa, Carena:2004xs, Schael:2013ita, Heeck:2014zfa}, we find $\lambda_{\rm mix} \lesssim 10^{-4}$ and the smallness of $\lambda_{\rm mix}$ is justified.  

After the breaking of the electroweak and the $B-L$ gauge symmetries, the  Dirac and Majorana masses for the neutrinos are generated in Eq.~(\ref{eq:seesaw0}):
\bea  
   {\cal L} \supset   \left(-\sum_{i,j=1}^{3} m_D^{ij} \overline{N_R^i} \; \nu_L^j  - \frac{1}{2} \sum_{i=1}^{3} M_{N^i} \overline{N_{R}^{i~C}} N^i_{R} \right) +{\rm h.c.},
\eea 
The resulting neutrino mass matrix is given by 
\bea
M_\nu = \begin{pmatrix}
0 & m_D^T \\
m_D & M_N 
\end{pmatrix}, 
\eea
which, for $M_N\gg m_D$, leads to the seesaw formula for the light neutrino mass matrix 
\bea
m_\nu \simeq m_D^T M_N^{-1} m_D. 
\label{eq:seesaw}
\eea
Using the Maki-Nakagawa-Sakata (MNS) matrix $U_{MNS}$ we can diagonalize $m\nu$ as 
\bea
D_\nu = U_{MNS}^T \; m_\nu \; U_{MNS},  
\eea
where $D_\nu = {\rm diag} (m_1, m_2, m_3)$ with the mass eigenvalues $m_i$. The explict form of the MNS matrix is given by 
\bea
U_{\rm MNS}
=
\left(\begin{array}{ccc}c_{12}c_{13} & s_{12}c_{13} & s_{13}e^{-i\delta} \\-s_{12}c_{23}-c_{12}s_{23}s_{13}e^{i\delta} & c_{12}c_{23}-s_{12}s_{23}s_{13}e^{i\delta} & s_{23}c_{13} \\s_{12}s_{23}-c_{12}c_{23}s_{13}e^{i\delta} & -c_{12}s_{23}-s_{12}c_{23}s_{13}e^{i\delta} & c_{23}c_{13}\end{array}\right)
\left(\begin{array}{ccc}e^{i\sigma_1} & 0 & 0 \\0 & e^{i\sigma_2} & 0 \\0 & 0 & 1\end{array}\right),
\eea
where
$s_{ij}=\sin\theta_{ij}$, $c_{ij}=\cos\theta_{ij}$, and
$\delta$ and $\sigma_i$ are the CP-violating Dirac and Majorana phases, respectively.
Neutrino mass parameters are determined by the oscillation data \cite{ParticleDataGroup:2022pth}: 
\bea
&& \sin^2 \theta_{12}= 0.307, \quad 
\sin^2 \theta_{13}= 2.20 \times 10^{-2}, \quad 
\Delta m_{21}^2 = 7.53 \times 10^{-5}\;  {\rm eV}, \quad 
\delta = 1.23 \pi \; {\rm rad} \nonumber \\
&& \sin^2 \theta_{23}= 0.534\; {\rm (NH)}, \quad  
\Delta m_{32}^2\equiv (m_3^2-m_2^2)= 2.437\times 10^{-3} \;  {\rm eV^2}\; {\rm (NH)}, \nonumber \\
&&\sin^2 \theta_{23}= 0.547\; {\rm (IH)}, \quad \; \Delta m_{32}^2\equiv (m_2^2-m_3^2)= 2.519\times 10^{-3} \;  {\rm eV^2}\; {\rm (IH)}, \qquad 
\label{eqn:oscidata}
\eea
Here, the Normal Hierarchy (NH) implies  $m_{lightest}\equiv m_1< m_2 <m_3$ while in the Inverted Hierarchy (IH) $m_{lightest}\equiv m_3< m_1 <m_2$. Hence, for the NH,    
\bea
m_2=\sqrt{m_{lightest}^2 +\Delta m_{12}^2 } ,\quad
m_3=\sqrt{m_{lightest}^2 +\Delta m_{12}^2+\Delta m_{23}^2} ,
\eea
while for the IH, 
\bea
m_1=\sqrt{m_{lightest}^2+ \Delta m_{23}^2-\Delta m_{12}^2 } ,\quad
m_2=\sqrt{m_{lightest}^2 +\Delta m_{23}^2 }. 
\eea
The neutrino mass matrix can be conveniently parametrized as \cite{Casas:2001sr,Ibarra:2003up}
\bea
m_D = \sqrt{M_N} O \sqrt{D_\nu} U_{\rm MNS}^\dag,
\label{eqn:CIR}
\eea
where $\sqrt{M_N}\equiv \mbox{diag}(\sqrt{m_{N^1}},\sqrt{m_{N^2}},\sqrt{m_{N^3}})$ and  
\bea
\sqrt{D_\nu} =\left\{
\begin{array}{ccc}
\sqrt{D_\nu^{\rm NH}} &=& {\rm diag} \left(\sqrt{m_{lightest}}, \sqrt{m_2}, \sqrt{m_3}\right),\\
\sqrt{D_\nu^{\rm IH}} &=& {\rm diag} \left(\sqrt{m_{lightest}}, \sqrt{m_1}, \sqrt{m_2}\right).
\end{array}\right.
\eea
In (\ref{eqn:CIR}), $O$ is a $3\times 3$ orthogonal matrix
\bea
O
=
\left(\begin{array}{ccc}C_{12}C_{13} & S_{12}C_{13} & S_{13} \\-S_{12}C_{23}-C_{12}S_{23}S_{13} & C_{12}C_{23}-S_{12}S_{23}S_{13}& S_{23}C_{13} \\S_{12}S_{23}-C_{12}C_{23}S_{13} & -C_{12}S_{23}-S_{12}C_{23}S_{13} & C_{23}C_{13}\end{array}\right),
\eea
where
$S_{ij}=\sin\omega_{ij}$ and $c_{ij}=\cos\omega_{ij}$ with complex numbers $\omega_{ij}$.

\section{Double Inflation in the Classically Conformal Minimal \texorpdfstring{$B-L$}{BL} Model}

Let us consider the following action in the Jordan frame with a non-minimal gravitational coupling between the inflaton ($\phi$) and the Ricci scalar curvature ($\cal R$) and  non-minimal kinetic term for $\phi$: 
\begin{eqnarray}
 {\cal S}_J &=& \int d^4 x \sqrt{-g} 
   \left[-\frac{1}{2} f(\phi)  {\cal R}+ \frac{1}{2} K (\phi) g^{\mu \nu} \left(\partial_\mu \phi \right) \left(\partial_\nu \phi \right) 
 - V_J (\phi) \right].  
\label{S_J}
\end{eqnarray}
Following Ref.~\cite{Karam:2023haj}, we set $f(\phi) = 1+ \xi \phi^2$ with $\xi>0$ being the non-minimal coupling parameter, the non-minimal kinetic term $K(\phi) =(1- \phi^2 / 6 \alpha)^{-n}$, and the inflaton potential $V_J (\phi)$ given by 
\begin{eqnarray}
   V_J(\phi) =  \frac{\lambda}{4} \left(  \frac{1}{2}\phi^4 \left(  \ln \left[ \frac{\phi^2}{v_\phi^2} \right] - \frac{1}{2} \right) + \frac{1}{4} v_\phi^4 \right). 
\label{eq:CW_potential1} 
\end{eqnarray}
This potential is essentially the same as the potential in Eq.~(\ref{eq:CW_potential}) with the identification of $\lambda = \frac{6}{11}\lambda_\Phi $ and $ \lambda = \beta_\Phi$. 
Throughout this paper we use the Planck unit by setting the reduced Planck mass $M_P=2.44 \times 10^{18}$ GeV to be 1.

After performing the conformal  transformation, $ g_{{\mu \nu}}\to g_{E {\mu \nu}} = f(\phi) g_{\mu \nu} $, the action in the Einstein frame is given by 
\begin{eqnarray}
S_E &=& \int d^4 x \sqrt{-g_E}\left[-\frac{1}{2}  {\cal R}_E + 
g_E^{\mu \nu} \left(\partial_\mu \sigma \right) \left(\partial_\nu \sigma \right)
   - V_E(\sigma) \right].  
\label{S_E}   
\end{eqnarray}
Here, $V_E$ is the potential in the Einstein Frame,  
\begin{eqnarray}
V_E (\phi(\sigma)) = \frac{V_J(\phi(\sigma))}{f(\phi(\sigma))^2}, 
\end{eqnarray}
and $\sigma$ (inflaton field in Einstein Frame) is related to the original field  $\phi$ by  
\begin{eqnarray}
 \left(\frac{d\sigma}{d\phi}\right)^{2} = \left( \frac{K}{f} + \frac{3}{2}\left(\frac{f^\prime}{f}\right)^2\right), 
\end{eqnarray}
where a ``$\it prime$" denotes a derivative with respect to $\phi$.  
Using this relation, the slow-roll parameters can be expressed in terms of $\phi$ as follows: 
\begin{eqnarray}
 \epsilon(\phi) &=& \frac{1}{2} \left(\frac{V_E^\prime}{V_E \; \sigma^\prime}\right)^2,   \nonumber \\
 \eta(\phi) &=& \frac{V_E^{\prime \prime}}{V_E \; (\sigma^\prime)^2}- \frac{V_E^\prime \; \sigma^{\prime \prime}}{V_E \; (\sigma^\prime)^3} ,   \nonumber \\
 \zeta (\phi) &=&  \left(\frac{V_E^\prime}{V_E \; \sigma^\prime}\right) 
 \left( \frac{V_E'''}{V_E \; (\sigma^\prime)^3}
-3 \frac{V_E'' \; \sigma''}{V_E \; (\sigma^\prime)^4} 
+ 3 \frac{V_E^\prime \; (\sigma^{\prime \prime})^2}{V_E \; (\sigma^\prime)^5} 
- \frac{V_E^\prime \; \sigma'''}{V_E \; (\sigma')^4} \right).  
\end{eqnarray}
The amplitude of the scalar curvature perturbation $\Delta_{\cal R}$ is given by 
\begin{equation} 
  \Delta_{\cal R}^2 = \left. \frac{V_E}{24 \pi^2 \epsilon } \right|_{k}, 
  \label{eq:scs}
\end{equation} 
where $k$ is the pivot scale. The Planck result sets the power spectrum $\Delta_\mathcal{R}^2= 2.195\times10^{-9}$ 
\cite{Planck:2018jri} for $k=0.05~\rm Mpc^{-1}$. 
The number of e-folds is given by
\begin{eqnarray}
  N_k = \frac{1}{\sqrt{2}} \int_{\phi_{\rm e}}^{\phi_k}
  d \phi  \frac{\sigma^\prime}{\sqrt{\epsilon(\phi)}}
\end{eqnarray} 
where $\phi_k$ is the inflaton value at horizon exit of the scale corresponding to $k$, and $\phi_e$ is the inflaton value at the end of inflation, which is defined by $\epsilon(\phi_e)=1$. To solve the horizon and flatness problems, $N_k=50-60$ is necessary.

The predictions for the inflationary observables, namely, the scalar spectral index $n_{s}$ and the tensor-to-scalar ratio $r$ are expressed in terms of slow-roll parameters as 
\begin{eqnarray}
n_s = 1-6\epsilon+2\eta, \; \;  \;\;\; 
r = 16 \epsilon,  \; \;  
\end{eqnarray} 
which are evaluated at $\phi=\phi_k$.

In our analysis, we consider the roll-back scenario discussed in Ref.~\cite{Karam:2023haj}. 
In this case, after the conclusion of the slow-roll inflation, which occurs at $\phi>v_\phi$, the inflaton rolls down towards a  potential minimum at $\phi = v_\phi$ and passes through the potential minimum, and then climbs up the potential hill towards the origin $\phi = 0$. 
The inflaton comes to a stop in the vicinity of $\phi = 0$, rolls back towards $\phi = v_\phi$, and then oscillates around the minimum until it decays to reheat the universe.  There are two phases of inflation  along with the evolution of inflaton. 
The slow-roll inflation takes place for $\phi>v_{\phi}$ which generates comsological perturbations corresponding to the observed CMB anisotropy. 
The second inflation occurs during the rolling back near $\phi = 0$. Since the curvature of the potential at $\phi = 0$ is zero in the CW potential, the ultra-slow roll condition is satisfied, and density fluctuations are significantly enhanced at small scales to form PBHs. 

It has been shown in Ref.~\cite{Karam:2023haj} that with a suitable choice of the parameters, namely, $\xi$, $n$, $\alpha$,  $\lambda$, $v_\phi$, the double-inflation can successfully produce inflationary predictions consistent with the CMB data and also generate large fluctuations to form PBHs. 
For our analysis, we choose the benchmark for the right panel of Fig.~2 in the paper~\cite{Karam:2023haj}: $\lambda = 1.79222\times 10^{-8}$ and $v_\phi = 0.3854781$, $n = 4$, and $\alpha =2v_\phi^2/3$. 
According to the paper, this benchmark predicts\footnote{Although the paper~\cite{Karam:2023haj} does not provide the explicit value for the tensor-to-scalar ratio ($r$) for this benchmark, Table~1 showing the results for various benchmark parameters indicates that the resultant inflationary predictions for $\xi \ll 1$ are almost the same, $r\simeq 0.016$.} $n_s=0.96072$ and for the number of e-folds $N_{0}=N_I + N_{II}= 55$, where $N_I = 41$ and $N_{II} = 14$ are the number of e-folds generated during the first and second phases of inflation, respectively. 
We note that $\phi_k \simeq \sqrt{6 \alpha} =2 v_\phi$, which is close to the singular point of the non-minimal kinetic function $K(\phi)$.

\section{Reheating after Inflation}

In the scenario we consider, after the end of the second inflation at $t_{end}$, the inflaton oscillates around the potential minimum at $\phi=v_\phi$ until it decays to reheat the universe. We use the sudden decay approximation, and all the inflation energy is transmitted to the radiation at $t = \tau_{\phi} = 1/\Gamma_\phi$, where $\tau_{\phi}$ is the lifetime of the inflaton and $\Gamma_\phi$ is its decay width. 
We assume that the universe is instantly thermalized at $t = \tau_{\phi}$.
This is the beginning of the Standard Big Bang cosmology. This radiation-dominated era continues until the matter-radiation equality is reached
at time $t_{eq}$. After that, the Universe stays matter-dominated until today $t_0$.

We now evaluate the number of e-folds based on the above picture. 
Denoting the co-moving wave number of the CMB scale by $k$, the scale factor by $a_k$ and the
Hubble parameter by $H_k$, they are related by $k = a_k H_k$ at the horizon exit scale. It follows that 
\begin{equation}
	\frac{k}{a_0 H_0}=\frac{a_k H_k}{a_0 H_0} = \frac{a_k}{a_{end}} \frac{a_{end}}{a_{rh}} \frac{a_{rh}}{a_{eq}}\frac{a_{eq}}{a_{0}} \frac{H_k}{H_0},
	\label{eq:k=aH1}
\end{equation}
where $a_{end}$, $a_{rh}$, $a_{eq}$, and $a_{0}$ are the scale factors at the end of inflation ($t_{end}$), at the reheating time ($\tau_\phi$), at matter-radiation equality ($t_{eq}$), and at present ($t_0$).   
Also, $H_0 = 100 h$ km\,s$^{-1}$ Mpc$^{-1}$ with $h= 0.674$ \cite{Planck:2018jri}. 
Each ratio in the right hand side of Eq.~(\ref{eq:k=aH1}) is evaluated as follows:
$\frac{a_k}{a_{end}} = e^{N_k}$;
$\frac{a_{end}}{a_{rh}} = \left(\frac{\rho_{rh}}{\rho_{end}}\right)^{1/3}$, where $\rho_{end}\simeq V_E (0)$ and $\rho_{rh} \simeq 3 \Gamma_\phi^2$ are the energy densities of inflaton at $t_{end}$ and $\tau_\phi$, respectively, and we have used the fact that the during the oscillation between $t_{end}$ and $\tau_{\phi}$ the inflaton potential is approximately quadratic, hence the equation-of-state $w=0$; 
$\frac{a_{rh}}{a_{eq}} = \left(\frac{\rho_{eq}}{\rho_{rh}}\right)^{1/4} \left(g_*^{eq}/g_*^{rh}\right)^{1/12}$, where $g_*^{eq} \simeq 43/11$, $g_*^{rh} \simeq 100$, $\rho_{eq} = 1.3 \times 10^{-110}$ is the energy density at the matter-radiation equality ($t_{eq}$), and we have used  $w=1/3$ for radiation; 
$a_{eq}= 1/(1 + z_{eq})$, where $z_{eq} = 3387$ \cite{ParticleDataGroup:2022pth} is the redshift of the matter-radiation equality; $H_k \simeq \sqrt{V_E (\phi_k)/3}$, where $V_E (\phi_k)$ is the value of the inflaton potential evaluated at the pivot scale. 
Plugging all the ratios we obtain the formula for e-folding number $N_k$:
\begin{equation}
	N_k \equiv \ln\frac{a_{end}}{a_k} \simeq 66.9 -\ln \frac{k}{a_0 H_0} +\frac{1}{12}\ln \frac{\rho_{rh}}{\rho_{end}} +\frac{1}{4}\ln \frac{V_E (\phi_k)}{\rho_{end}}+\frac{1}{4}\ln V_E (\phi_k) + \frac{1}{12} \left(\ln g_*^{eq} - \ln g_*^{rh}\right). 
	\label{eq:k=aH}
\end{equation}
For the fixed value of $N_k =55$ and the pivot scale $k$, this equation determines $\Gamma_\phi$.

\begin{figure}
	\centering
	\includegraphics[width=.6\textwidth]{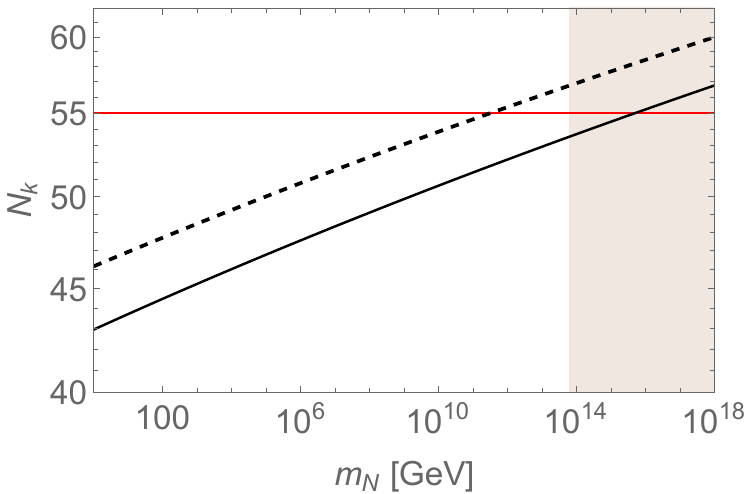}
	\caption{
 The plot shows $N_k$ as a function of $m_{N}$ for 
 two pivot scales, $k =0.05$ (diagonal solid line) and $k=0.002$ (diagonal dashed line). 
The gray shaded region depicts $m_{N} > m_\phi/2$, which is excluded by kinematics.
}
	\label{fig:reheat}
\end{figure}

Let us discuss the decay modes for inflaton in the classically conformal $U(1)_{B-L}$ model. 
Since the inflaton is identified with the $B-L$ Higgs boson, there are two possible decay modes: inflaton decays to SM Higgs pair through the mixed quartic coupling ($\lambda_{mix}$) in Eq.~(\ref{Higgs_Potential}) and inflaton decays to a pair of right-handed neutrinos through the Majorana Yukawa couplings ($Y_M^i$) in Eq.~(\ref{eq:seesaw}). 
The partial decay width for $\phi \to H^\dagger H$ is given by
\bea
\Gamma_\phi (\phi\to H^\dagger H) \simeq  \frac{\lambda_{\rm mix}^2 v_\phi^2}{8\pi m_\phi} = \frac{m_h^4}{8 \pi {\sqrt \lambda } v_\phi^3}, 
\label{eq:phiHH}
\eea
where we have used $m_\phi^2 = \lambda v_\phi^2$ and $\lambda_{\rm mix} = m_h^2/v_\phi^2$.  
The partial decay width for $\phi \to N_R^i N_R^i$ is given by
\bea
\Gamma_\phi (\phi\to N_R^i N_R^i) \simeq  \frac{(Y_{M}^i)^2}{16\pi} m_\phi  = \frac{\sqrt{\lambda} m_{N^i}^2}{8\pi v_\phi}, 
\label{eq:phiNN}
\eea
where we have used $m_{N^i} = Y_{M}^i v_\phi/\sqrt{2}$ and assumed $m_{N^i} < m_\phi/2$ with $m_\phi \simeq 1.3 \times 10^{14}$ GeV for our benchmark\footnote{Note that the vacuum stability condition, $m_{Z^\prime} >  \left(\frac{2}{3}\right)^{1/4} \sum_{i=1}^{3} m_{N^i}$, sets a lower bound on $Z^\prime$ boson mass, $m_{Z^\prime} \gtrsim 10^{11}$ GeV.}. 
For a successful thermal leptogenesis we will discuss below, the lightest right-handed neutrino mass $m_{N^1}\gtrsim 1$ TeV. 
We have assumed that  reheating occurs after the second inflation. 
This assumption leads to an upper bound on the decay width, $ \Gamma_{\phi}^{\rm max } < \sqrt{H(\phi \simeq 0)} = \sqrt{\frac{V_E (0)}{3}}$, which is satisfied for $m_{N^i} < m_\phi/2$.

Assuming the inflaton total decay width is dominated by its decay mode to a single generation of right-handed neutrino pair, Eq.~(\ref{eq:k=aH}) relates $N_k$ with $m_{N^i}$ for a fixed pivot scale $k$. 
In Fig.~\ref{fig:reheat}, we show  $N_k$ as a function of $m_{N^i}$ for $k =0.05$ (diagonal solid line) and $k=0.002$ (diagonal dashed line). 
The gray shaded region depicting $m_{N^i} > m_\phi/2$ is excluded. 
For the benchmark 
value $N_k=55$ (the horizontal red line), we have found a solution $m_{N^i} = 3.1 \times 10^{11}$ GeV only for $k = 0.002$, which corresponds to $\Gamma_\phi \simeq 0.55$ GeV. 
The corresponding reheating temperature is estimated by $T_R \simeq \left(\frac{90}{\pi^2 g_*}\right)^{1/4}\sqrt { \Gamma_\phi (\phi \to N_R^i N_R^i)}$ which yields $T_R = 6.4\times 10^8$ GeV. 
Although $m_{N^i} > T_R$, this estimate is justified if $N_R^i$ decay width is greater than the inflaton decay width, and created $N_R^i$'s instantly decay to the SM particles. 
Here we calculate the partial decay width of $N_R^i$ to the SM final states, $\Gamma(N_R^i\to H \ell_L^j)$, and its total total decay width is found to be
\bea
\Gamma_i = \Sigma_j \Gamma(N_R^i\rightarrow\ell_L^j+H) = \frac{m_{N^i}}{8\pi} (Y_D Y_D^\dagger)_{ii}  \; = \frac{m_{N^i}^2}{4\pi v_h^2} \left( O D_\nu O^\dagger \right)_{ii}  \simeq \frac{1}{4\pi} \frac{m_{N_i}^2 m_i}{v_h^2},
\label{eq:NSMSM}
\eea
where we have applied the seesaw formula Eq.~(\ref{eq:seesaw}) and assumed $O\simeq {\bf 1}_{3 \times 3}$.  
Using $m_{N^i} =  3.1\times 10^{11}$ GeV, we find that $\Gamma_i/{\rm GeV} \simeq 1.3\times 10^7 \left(\frac{m_i}{ 0.1\; {\rm eV}}\right)\; {}\gg \Gamma_\phi/{\rm GeV}\simeq 0.55$.

\section{Leptogenesis}

We now consider the leptogenesis scenario from out-of-equilibrium $N$ decay which generates lepton asymmetry. The lepton asymmetry is then converted to baryon asymmetry by the so-called Sphaleron process \cite{Manton:1983nd,Klinkhamer:1984di,Arnold:1987mh}. 
The observed baryon number asymmetry $n_B/s \simeq 10^{-10}$ \cite{Planck:2018jri,ParticleDataGroup:2022pth} originates from the lepton number asymmetry, 
\bea
\frac{n_B}{s} \simeq \frac{n_L}{s}, 
\eea
where $n_{B}$ ($n_{L}$) is the baryon (lepton) number density, and $s$ is the entropy density. 
In the leptogensis, $n_L$ is controlled by the CP-asymmetry parameter $\varepsilon$ \cite{Covi:1996wh}:
\bea
\varepsilon_i&\equiv&\frac{
\sum_j\left[\Gamma (N_R^i\rightarrow\ell_L^j+H)
-\Gamma(N_R^i\rightarrow\overline {\ell_L^j}+H^\dag)\right]}{\sum_j\left[\Gamma (N_R^i\rightarrow\ell_L^j+H)
+\Gamma(N_R^i\rightarrow\overline {\ell_L^j}+H^\dag)\right]}\nonumber \\
&=&
-\sum_{j\neq i}
\frac{{\rm Im}[(m_D m_D^\dag)^2_{ij}]}{(m_Dm_D^\dag)_{ii}(m_Dm_D^\dag)_{jj}}
\frac{m_{N^i}}{m_{N^j}} \frac{ \Gamma_j}{m_{N^j}}\left( \frac{1}{2}V_j+S_j\right),
\label{eqn:epsilon}
\eea
where 
\bea
V_j&=& 2\frac{m_{N^j}^2}{m_{N^i}^2}\left[\left(1+\frac{m_{N^j}^2}{m_{N^i}^2}\right)\ln\left(1+\frac{m_{N^i}^2}{m_{N^j}^2}\right)-1\right],\\
S_j&=&\frac{m_{N^j}^2\Delta M_{ji}^2}{(\Delta M_{ji}^2)^2+m_{N^i}^2\Gamma_j^2},
\eea
are contributions from the decay including the vertex corrections and self-energy corrections, respectively, with $\Delta M_{ji}^2=m_{N^j}^2-m_{N^i}^2$. 
We can see that $ V_j\simeq S_j\simeq 1$ if the right-handed neutrino mass spectrum is hierarchical. 
If two right-handed neutrinos $N_i$ and $N_j$ are degenerate in mass but $\Delta M_{ji}^2 \gtrsim m_{N^i} \Gamma_j$, then $S_j \simeq m_{N^j}^2 /\Delta M_{ji}^2 
 = \frac{1}{1- (m_{N^j}/m_{N^i})^2}\gg 1$ is enhanced. 
By using the Eq.~(\ref{eqn:CIR}), we rewrite  Eq.~(\ref{eqn:epsilon}) as 
\bea
\varepsilon_i=
-\sum_{j\neq i}
\frac{{\rm Im}[(O D_\nu O^\dagger)^2_{ij}]}{(O D_\nu O^\dagger)_{ii}} \frac{\left( \frac{1}{2}V_j+S_j\right)}{4\pi v_h^2}\; m_{N^i}. 
\label{eqn:epsilon1}
\eea
Note that $\varepsilon_i$ is independent of the MNS matrix and is determined 
by the orthogonal matrix $O$, and both light and heavy neutrino mass spectra.

There are two possibilities depending on whether (i) the lightest right-handed neutrino is thermalized in the SM particle plasma  or (ii) it has never been thermalized. 
In case (i), there is a lower bound on the lightest right-handed neutrino mass $m_{N^1} > 10^{9-10}$ GeV \cite{Buchmuller:2002rq, Buchmuller:2004nz} for the successful leptogenesis if the right-handed neutrino mass spectrum is hierarchical. 
In thermal leptogenesis, if the washing-out effects are negligible, the resultant baryon-to-entropy ration is given by $n_B/s \sim n_L/s \sim \varepsilon_1/g_*  \simeq 0.01 \times \varepsilon_1$, which gives a lower bound on $\varepsilon_1\gtrsim 10^{-8}$ to reproduce the observed value $n_B/s \simeq 10^{-10}$. This condition corresponds to the lower bound on $m_{N^1} > 10^{9-10}$ GeV. 
Since $N_R^1$ is thermalized before it decays, $T_R \gtrsim m_{N^1}$, the lower bound on $m_{N^1} > 10^{9-10}$ GeV is not satisfied in our case. 
However, if the lighter of the two right-handed neutrinos are almost degenerate in mass, $S_j$ is enhanced as we have discussed before,  and as a result the CP asymmetry parameter $\varepsilon$ is enhanced. 
Hence, a successful leptogenesis can still be realized even for $m_{N^1} < 10^{9-10}$ GeV. 
This scenario is known as resonant leptogenesis \cite{Flanz:1996fb,Pilaftsis:1997jf}. 
We can apply resonant leptogenesis by setting  $m_{N^1}\simeq m_{N^2} < T_R = 6.4 \times 10^8$ GeV and identifying  $m_{N^3} = m_N = 3.1\times 10^{11}$ GeV with the heaviest right-handed neutrino mass.

In case (ii), we can identify $m_{N^1} = m_N= 3.1\times 10^{11} $ GeV with the lightest right-handed neutrino while we set $m_{N^{2,3}} > m_\phi/2$. Since $m_{N^1}> T_R$, $N_R^1$ has never been in thermal equilibrium with the SM plasma. 
The universe is thermalized by $N_R^1$ decay to SM particles as discussed before, and this decay also generates lepton asymmetry of the universe. 
This scenario is called non-thermal leptogenesis \cite{Lazarides:1990huy,Kumekawa:1994gx}. 
The generated lepton number $n_L$ is estimated as 
\bea
n_{L} \sim n_\phi \times \varepsilon, 
\eea
where $n_\phi \sim \frac{\rho_{rh}}{m_\phi}$ is the number density of inflaton when it decays. 
Using the entropy density $s \sim \frac{\rho_{rh}}{T_R}$, we obtain $\frac{n_B}{s}\sim \frac{T_R}{m_\phi} \; \varepsilon = 5.1\times 10^{-6}\; \varepsilon$. 
Note that since $N_R^1$ is never in thermal equilibrium, there is no washing-out effect.   
To reproduce the observed baryon asymmetry, $n_B/s \simeq 10^{-10}$, we adjust $\varepsilon \sim 10^{-5}$. 
For the hierarchical right-handed neutrino mass spectrum,  
\bea
\varepsilon_1\simeq 
- \frac{3 m_{N^1}}{8\pi v_h^2}\; \sum_{j=2}^3
\frac{{\rm Im}[(O D_\nu O^\dagger)^2_{1j}]}{(O D_\nu O^\dagger)_{11}}   . 
\eea
For example, we find $\varepsilon_1  = 10^{-5}$ with $m_{lightest} \simeq 10^{-3}$ eV is satisfied for NH  with $\omega_{12} \simeq +1.85 \; i $, 
$\omega_{23}=0$ and $\omega_{13}= \frac{\pi}{4} -2\; i $, while $\omega_{12} \simeq +0.87 \; i $, $\omega_{23}=0$ and $\omega_{13}= \frac{\pi}{4} +2\; i$ for IH.

\section{Conclusions and Discussions}

The classically conformal minimal $B-L$ model is particularly interesting because it can address 
(i) the origin of the electroweak symmetry breaking triggered by the radiative $B-L$ symmetry breaking with the Coleman-Weinberg potential; (ii) the origin of the neutrino mass through the seesaw mechanism with the heavy right-handed Majorana neutrinos whose masses are generated by the $B-L$ symmetry breaking; 
(iii) the origin of the baryon-asymmetry of the universe through leptogenesis with the CP-asymmetric out-of-equilibrium decay of the heavy neutrinos.

It has recently been shown in Ref.~\cite{Karam:2023haj} that the double-inflation scenario based on the Coleman-Weinberg potential can successfully generate PBHs, which can be a dark matter candidate of the universe, while the inflationary predictions are consistent with the Planck measurements. 
In this paper, we have proposed a UV completion of the scenario discussed in Ref.~\cite{Karam:2023haj} 
by the classically conformal minimal $B-L$ model with the identification of the $B-L$ Higgs field to be the inflaton. 
We have shown this UV completion leads to a viable history of the early universe after the double inflaton: the universe is reheated via inflaton decay into a pair of right-handed neutrinos, and 
the reheating temperature is determined by its relation to the number of e-folds, and then the right-handed neutrino mass is fixed. 
With the determined values of reheating temperature and right-handed neutrino mass, we have considered 
baryogenesis via leptogenesis, and found that either thermal or non-thermal leptogenesis can successfully reproduce the observed baryon asymmetry in the universe consistently with the neutrino oscillation data.

\begin{figure}
	\centering
\includegraphics[width=.7\textwidth]{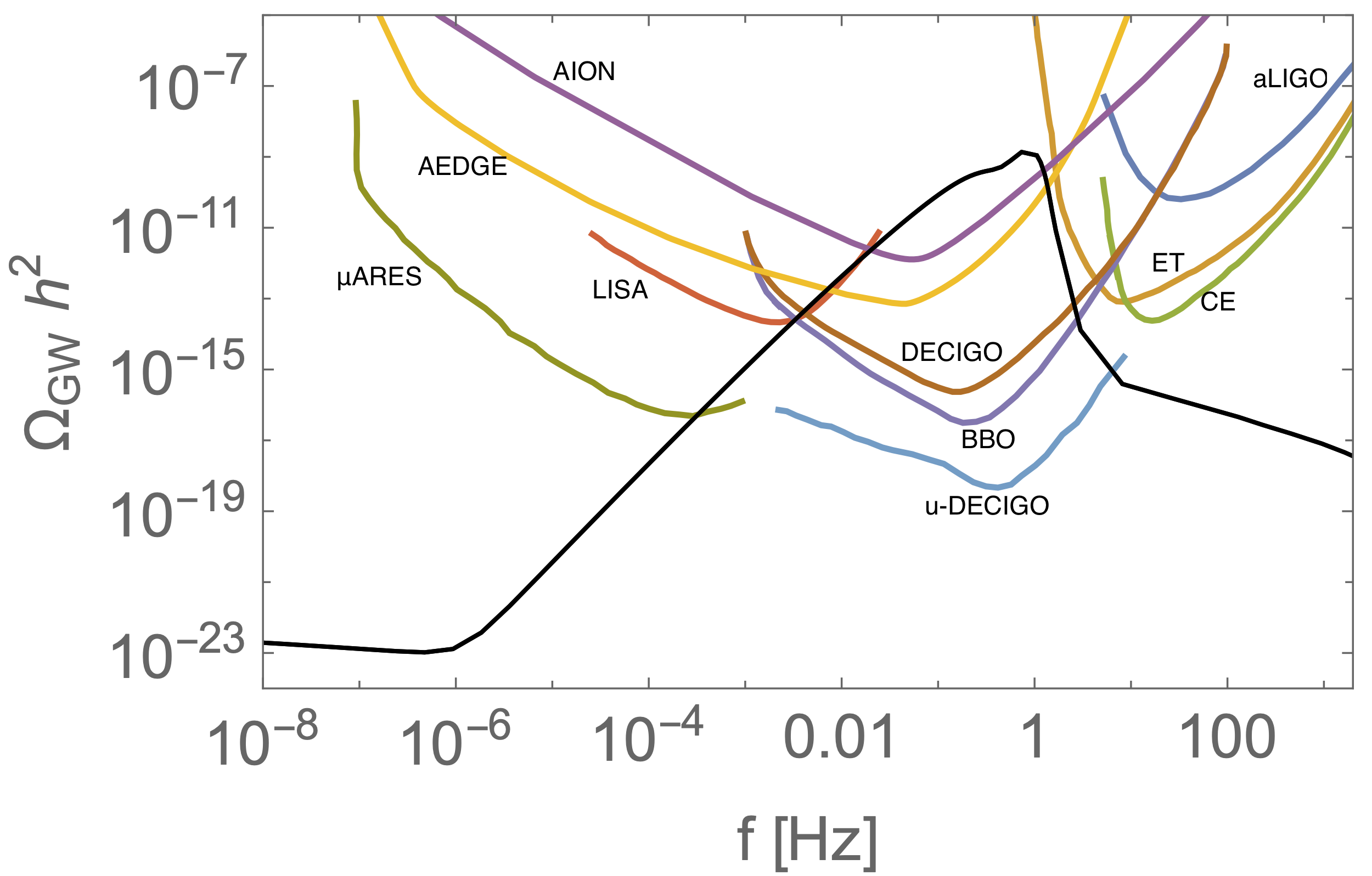}
	\caption{The scalar-induced GW spectrum obtained by using the power spectrum shown in Fig.~2 of Ref.~\cite{Karam:2023haj}, along with the current and future reach of various GW observations.}
	\label{fig:GW}
\end{figure}

Finally, we discuss the scalar-induced GWs from the enhanced power spectrum associated with PBH production. 
Although in the first-order perturbation, the scalar and tensor modes do not mix, the scalar modes can source tensor modes in the second order of perturbation \cite{Matarrese:1993zf,Matarrese:1997ay,Nakamura:2004rm,Ananda:2006af,Baumann:2007zm} (see Ref.~\cite{Domenech:2021ztg} for a review).  
As shown in Fig.~2 of Ref.~\cite{Karam:2023haj}, 
the second inflation remarkably enhances the scalar perturbations and the power spectrum (${\cal P}_\zeta$)\footnote{$\Delta_{\cal R}^2 $ in Eq.~(\ref{eq:scs}) is defined as ${\cal P}_\zeta$ evaluated at the pivot scale $k$.} 
exhibits a sharp peak at $k_{\rm peak} = {\cal O (}10^{14})$ Mpc$^{-1}$. 
The temperature of the universe when this scale re-entered the horizon scale ($T_{ren}$) is estimated 
as $T_{ren} \sim \frac{k_{\rm peak}}{2 \pi}\frac{M_P}{T_0} \sim 10^6$ GeV, where $T_0 \simeq 2.7$ K
is the present CMB temperature. 
Hence, the fluctuations re-entered the horizon during the radiation-dominated era 
after the reheating at $T_R =6.4 \times 10^8$ GeV. 
In this case, the scalar-induced GW spectrum is calculated by \cite{Kohri:2018awv} 
\bea \label{Omega_GW}
\Omega_{\rm GW} h^2 \approx 2.14\times 10^{-3} 
\frac{g_*}{106.75}
\left(\frac{g_{*,s}}{106.75}\right)^{-\frac{4}{3}}
\!\!\int_{-1}^1 \!\!{\rm d} x \!\!\int_1^\infty \!\!{\rm d} y \, \mathcal{P}_\zeta\left(\frac{y-x}{2}k\right) \mathcal{P}_\zeta\left(\frac{x+y}{2}k\right) F(x,y) \,,
\label{scalar-induced}
\eea
where
\bea
F(x,y) &=& \frac{(x^2\!+\!y^2\!-\!6)^2(x^2-1)^2(y^2-1)^2}{(x-y)^8(x+y)^8} \\ \nonumber
&& \times  \left\{\left(x^2-y^2+\frac{x^2\!+\!y^2\!-\!6}{2}\ln\left|\frac{y^2-3}{x^2-3}\right|\right)^{\!2} + \frac{\pi^2(x^2\!+\!y^2\!-\!6)^2}{4}\theta(y-\sqrt{3}) \right\} .
\eea
The resultant power spectrum from the double inflation for our benchmark is shown in the right panel of Fig.~2 in Ref.~\cite{Karam:2023haj}. 
Employing their power spectrum as an input to Eq.~(\ref{scalar-induced}), we calculate the scalar-induced GW spectrum. 
Our result is shown in Fig.~\ref{fig:GW}. 
The sharp peak of $P_\zeta$ at $k_{\rm peak} = {\cal O (}10^{14})$ Mpc$^{-1}$ induces the remarkable enhancement of the GW spectrum at $f \sim 1$ Hz.  
Interestingly, the predicted scalar-induced GW spectrum well-overlaps with the search reach
of the proposed/planned GW observations, like LISA, DECIGO, BBO, SKA and ET~\cite{Audley:2017drz,Sato:2017dkf,Sathyaprakash:2009xs,Zhao:2013bba,Yagi:2011wg}.

\section*{Acknowledgement}
\label{Asck}
The work of N.O. is supported in part by the United States Department of Energy Grant 
No.~DE-SC0012447 and DE-SC0023713. AP thanks Indian Association for the Cultivation of Science, Kolkata for financial support through Research Associateship, where part of the work was done. AP also wishes to thank the Indo-French Centre for the Promotion of Advanced Research for supporting the postdoctoral fellowship through the proposal 6704-4 under the Collaborative Scientific Research Programme.

\printbibliography
\end{document}